\documentclass[onecolumn]{revtex4}
\usepackage{graphicx}% Include figure files
\usepackage{dcolumn}% Align table columns on decimal point
\usepackage{amsmath}
\usepackage{amssymb}
\usepackage{bm}% bold math
\usepackage{amsmath}
\usepackage{amsfonts}
\usepackage{amssymb}

\input epsf

\begin{document}

\title{\Large Is Emergent Universe a Consequence of Particle
Creation Process ?}

\author{\bf Subenoy Chakraborty
\footnote{schakraborty.math@gmail.com} }

\affiliation{Department of Mathematics,Jadavpur
University,Kolkata-700032,India}

\date{\today}

\begin{abstract}
A model of an emergent universe is formulated using the mechanism
of particle creation. Here the universe is considered as a
non-equilibrium thermodynamical system with dissipation due to
particle creation mechanism. The universe is chosen as spatially
flat FRW space-time and the cosmic substratum is chosen as perfect
fluid with barotropic equation of state.Both first and second
order deviations from equilibrium prescription is considered and
it is found that the scenario of emergent universe is possible in
both the cases.
\end{abstract}

\maketitle

\textbf{Keywords:}Emergent scenario,Particle Creation ,
Non-equilibrium Thermodynamics.

Pacs nos.: 98.80Cq,98.80.-k.

\section{Introduction}

        To overcome the initial singularity (big bang) of standard
        cosmology ,there are various proposed cosmological
        scenarios which can be classified as bouncing universes or
        the emergent universes. Here we shall focus on the second
        choice which arises due to the search for singularity free
        inflationary models in the context of classical general
        relativity. In fact an emergent universe is a model
        universe in which in which there is no time-like
        singularity, ever existing and having almost static
        behavior in the infinite past ($t\rightarrow -\infty$).
        Eventually the model evolves into an inflationary stage.
        Also the emergent universe scenario can be said to be a
        modern version and extension of the original Lemaitre-
        Eddington universe.\\

        Harrison [1] in 1967 obtained a model of the closed
        universe with only radiation and showed that
        asymptotically(as $t\rightarrow -\infty$)it approaches the
        state of an Einstein static model.Then after a long gap,
        Ellis and Maartens [2] ,Ellis et. al [3] in recent past were
        able to formulate closed universes with a minimally
        coupled scalar field $\phi$ with a special form for the
        self interacting potential and possibly some ordinary
        matter with equation of state $p = \omega \rho
        (-\frac{1}{3}\leq\omega\leq1)$. However, exact analytic
        solutions were not presented in their work,only the
        asymptotic behavior agrees with emergent universe
        scenario.Subsequently, Mukherjee etal [4] obtained
        solutions for Starobinsky model with features of an
        emergent universe. Also Mukherjee etal [5] formulated a
        general framework for an emergent universe model using an
        adhoc equation of state which has exotic behavior in some
        cases. Afterwords, a lot of works [6-13] has been done to
        model emergent universe for different gravity theories as
        well as for various type of matter . Very recently,the
        idea of quantum tunneling has been used to model emergent
        universe [14] . Here the initial static state is
        characterized by a scalar field in a false vacuum and then
        it decays to a state of true vacuum through quantum
        tunneling.\\
        From the thermodynamical aspect it has been proposed that
        entropy consideration favors the Einstein static state as
        the initial state for our universe [15,16]. Also
        recently, Pavan et. al [17,18] have examined the validity of
        the generalized second law of thermodynamics in the
        transition from a generic initial Einstein static phase to
        the inflationary phase and also from the end of the
        inflation to the conventional thermal radiation dominated
        era. In this context, the present work is quite different.
        Here universe is considered  as a non-equilibrium
        thermodynamical system with dissipative phenomena due to
        particle creation . Both first and second order deviations
        from equilibrium configuration is taken into account and
        emergent universe solutions are possible in both the
        cases. The paper is organized as follows : Section 2
        describes the particle creation in cosmology from the
        perspective of non-equilibrium thermodynamics . Emergent
        universe scenario has been presented both for first and
        second order non-equilibrium thermodynamics in sections 3
        and 4 respectively. Finally, summary of the present work
        has been presented in section 5.\\

\section{Particle Creation in Cosmology : Non-equilibrium
        Thermodynamics}

Suppose there are N particles in a closed thermodynamical system having internal energy E.
Then the first law of thermodynamics is essentially the conservation of internal energy as [19]

\begin{equation}
dE = dQ - p dV
\end{equation}

where as usual p is the thermodynamic pressure, V is any co-moving
volume and $dQ$ represents the heat received by the system in time
dt. By introducing, the energy density $\rho = \frac{E}{V}$, the
particle number density $n=\frac{N}{V}$ and heat per unit particle
$dq=\frac{dQ}{N}$, the above conservation equation can be
rewritten as

\begin{equation}
T ds=dq=d(\frac{\rho}{n}) + p d(\frac{1}{n})
\end{equation}

This is referred to as Gibbs equation with $s$, the entropy per
particle. This equation is also true when the particle number is
not conserved i.e. the system is not a closed system [20].\\
Thus
for an open thermodynamical system, the non-conservation of fluid
particles ($N^{\mu}_{;\mu}\neq 0$) is expressed mathematically as
\begin{equation}
\dot{n}+ \Theta n = n\Gamma
\end{equation}
where $N^{\mu} = n u^{\mu}$ is the particle flow vector,$u^{\mu}$
is the particle four velocity , $\Theta=u^{\mu}_{;\mu}$ is the
fluid expansion , $\Gamma$ is termed as the rate of change of the
particle number in a comoving volume V and by notation $\dot{n}=
n_{,\mu} u^{\mu}$. The positivity of the parameter $\Gamma$
indicates creation of particles while there is annihilation of
particles for $\Gamma<0$. Any non-zero $\Gamma$ will behave as an
effective bulk pressure of the thermodynamical fluid and
non-equlibrium thermodynamics should come into picture [21].\\
 We
shall consider spatially flat FRW model of the universe as an open
thermodynamical system which is non-equilibrium in nature due to
particle creation mechanism. Now the Einstein field equations are

\begin{equation}
\kappa \rho =3 H^{2} ,~~~ \kappa(\rho+p+\Pi)=-2\dot{H}
\end{equation}

where the cosmic fluid is characterized by the energy-momentum
tensor

\begin{equation}
T_{\mu}^{\nu}=(\rho + p + \Pi)u_{\mu}u^{\nu} + (p + \Pi)g_{\mu}^{\nu}
\end{equation}

The energy conservation relation $T^{\mu\nu}_{;\nu}=0$ takes the
form

\begin{equation}
\dot{\rho} + 3 H (\rho + p+ \Pi)=0
\end{equation}

In the above Einstein field equations (i.e. equation(4))$\kappa =
8 \Pi G$ is the Einstein's gravitational constant and the pressure
term $\Pi$ is related to some dissipative phenomena (say bulk
viscosity).\\
 However, in the present context , the cosmic fluid
may be considered as perfect fluid where the dissipative term
$\Pi$ is the effective bulk viscous pressure due to particle
creation or equivalently the conventional dissipative fluid is not
taken as cosmic substratum,rather a perfect fluid with varying
particle number is considered. This equivalence can be nicely
described for adiabatic (or isentropic) particle production as
follows [21-23].
Now ,using the conservation equations (3) and (6)
the entropy variation can be obtained from the Gibbs equation(2)
as

\begin{equation}
n T \dot{s} = - 3 H \Pi - \Gamma (\rho + p)
\end{equation}

with $T$ , the temperature of the fluid . If the thermodynamical
system is chosen as an adiabatic system i.e. entropy per particle
is constant (variable in dissipative process) ($\dot{s}$) , then
from the above relation (7) the effective bulk pressure is
determined by particle creation rate as

\begin{equation}
\Pi=-\frac{\Gamma}{3 H} (\rho + p)
\end{equation}

Thus for isentropic thermodynamical process a perfect fluid with
particle creation phenomena is equivalent to a dissipative fluid.
Further, it should be noted that in the adiabatic process the
entropy production is caused by the enlargement of the phase space
(also due to expansion of the universe in the present model). This
effective bulk pressure does not correspond to conventional
non-equilibrium phase , rather a state having equilibrium
properties as well (note that it is not the equilibrium era with
$\Gamma=0$).\\
 Elimination of the effective bulk pressure from the
Einstein field equations (4) using the isentropic condition (8) we
have

\begin{equation}
\frac{\Gamma}{3 H} = 1 + \frac{2}{3 \gamma}(\frac{\dot{H}}{H^{2}})
\end{equation}

with $\gamma$ , the adiabatic index (i.e.$p=(\gamma - 1)\rho$).
Thus , if we know the cosmological evolution then from the above
equation the particle creation rate can be determined or otherwise
,assuming the particle creation rate as a function of the Hubble
parameter one can determine the corresponding cosmological phase
which we shall try in the next two sections.

\section{ Emergent Universe  in first order non-equilibrium
thermodynamics}

    In first order theory due to Eckart[24] the entropy flow
    vector is defined as

\begin{equation}
 s_{E}^{\mu}= n s u^{\mu}
\end{equation}

So using the number conservation equation (3) and the isentropic
condition (8) we obtain (suffix stands for the corresponding
variable in Eckart's theory)

\begin{equation}
\left(s_{E}^{\mu}\right)_{;\mu} = - \frac{\Pi_{E}}{T}(3 H + \frac{n \mu \Gamma_{E}}{\Pi_{E}})
\end{equation}

where

\begin{equation}
\mu = (\frac{\rho +p}{n}) - T s
\end{equation}

is the chemical potential. As it has been shown above that
particle production is effectively equivalent to a viscous
pressure ,so for the validity of the second law of thermodynamics
i.e.$\left(s_{E}^{\mu}\right)_{;\mu}\geq 0$, it is reasonable to assume [22]

\begin{equation}
\Pi_{E} = - \zeta(3 H + \frac{n \mu \Gamma_{E}}{\Pi_{E}})
\end{equation}

As a result, we have

\begin{equation}
\left(s_{E}^{\mu}\right)_{;\mu} = \frac{\Pi_{E}^{2}}{T \zeta} \geq 0
\end{equation}

where $\zeta$ is termed as bulk viscous coefficient and the bulk
viscous pressure satisfies the inhomogeneous  quadratic relation

\begin{equation}
\Pi_{E}^{2} + 3 \zeta \Pi_{E} H = - \zeta n \mu \Gamma_{E}
\end{equation}

It should be noted that the familiar linear relation for bulk
viscous pressure i.e.$\Pi_{E}=-3 \zeta H$ may be recovered from
the above quadratic relation (15) either by $\Gamma_{E}=0$ or
$\mu=0$. Now using equation (8) in (13) to eliminate $\Gamma_{E}$
we have

\begin{equation}
\Pi_{E}=-3 \zeta_{eff} H
\end{equation}

where $\zeta_{eff} = (\frac{n s T}{\rho + p})\zeta$.\\
 The second
Friedmann equation in equation (4) and using equation (16) one
obtains the differential equation in H as

\begin{equation}
2 \dot{H} =-3 \gamma H^{2} + 3 \zeta \kappa H
\end{equation}

where for simplicity chemical potential is chosen to be zero so
that $\zeta_{eff} = \zeta$.Now solving equation (17) the Hubble
parameter can be obtained as

\begin{equation}
H = \frac{3 \zeta_{o}}{{3 \gamma+exp(-\tau)}}
\end{equation}

where $\tau= \frac{3 \zeta_{0}}{2}(t - t_{0})$, $\zeta_{0} = \zeta
\kappa$ and $t_{0}$ is the constant of integration. Integrating
once more , the scale factor has the solution

\begin{equation}
(\frac{a}{a_{0}})^{\frac{3 \gamma}{2}} = 3 \gamma exp(\tau) + 1
\end{equation}

with $a_{0} = a(t_{0})$.
We see that as $t\rightarrow -\infty$,
$a\rightarrow a_{0}$ and for $ t<< t_{0}$, $a\simeq a_{0}$, while
$a$ grows exponentially for $t> t_{0}$, Thus the above
cosmological solution has the following asymptotic  features:\\

i)$a\rightarrow a_{0}$, $H\rightarrow 0$ as $t\rightarrow
-\infty$.\\

ii)$a\simeq a_{0}$ ,$H\simeq 0 $ for $t<<t_{0}$.\\

iii) $a\simeq exp(H_{0} (t-t_{0}))$, $H_{0} =
\frac{\zeta_{0}}{\gamma}$ for
$t>>t_{o}$.\\

Therefore, in the first order Eckart theory ,the
particle production process gives rise to a cosmological solution
that describes a scenario of emergent universe.

\section{Emergent Universe Scenario from Second Order Non-equilibrium
Thermodynamics}

If we consider the second order deviations from equilibrium , then
according to Israel and Stewart the entropy flow vector is chosen
as[22,23]

\begin{equation}
S^{\mu} = [s n - \frac{\tau \Pi^{2}}{2 \zeta T}]u^{\mu}
\end{equation}

where $\tau$ is the relaxation time that usually describes the
relaxation from states with $\dot{s}>o$ to those with $\dot{s}=0$
due to internal (elastic) collision processes and is of the order
of the mean free collision time [21,22].The essential physical
difference between the non causal (first order theory) and the
present causal (second order) theory is the introduction of a
finite relaxation time. In fact, in a causal theory if the
particle production rate $\Gamma$ has been switched off then the
effective bulk viscous pressure will decay to zero during the
relaxation time $\tau$.(Note that in the present cosmological
context the relaxation time is of the order of Hubble time
$H^{-1}$).
Now proceeding as in the previous section,the
expression for the entropy production density is

\begin{equation}
S^{\mu}_{;\mu}=-\frac{\Pi}{T}[3 H+ \frac{\tau}{\zeta}\dot{\Pi} +\frac{1}{2}\Pi T(\frac{\tau}{\zeta T}u^{\mu})_{;\mu}]
\end{equation}

So for the validity of the second law of thermodynamics we choose

\begin{equation}
\Pi = -\zeta [3 H+ \frac{\tau}{\zeta}\dot{\Pi} +\frac{1}{2}\Pi T(\frac{\tau}{\zeta T}u^{\mu})_{;\mu}]
\end{equation}

and hence $S^{\mu}_{;\mu}= \frac{\Pi^{2}}{\zeta T}\geq 0$.\\
 As a
result ,the effective bulk viscous pressure now becomes a
dynamical variable with causal evolution equation [22,23]

\begin{equation}
\Pi + \tau \dot{\Pi} = -3 \zeta H - \frac{1}{2} \Pi \tau [3 H + \frac{\dot{\tau}}{\tau}-\frac{\dot{\zeta}}{\zeta}-\frac{\dot{T}}{T}].
\end{equation}

It should be noted that as $\tau\rightarrow 0$, the effective bulk
viscous pressure is no longer be a dynamical variable and it
satisfies the usual relation $\Pi=-3 \zeta H$.
However,due to
complicated form of the evolution equation,$\Pi$ can not be solved
.So for cosmological solution we choose $\Gamma$
phenomenologically and solve the differential equation in H (see
equation (9)). If for simplicity ,we assume the particle
production rate $\Gamma$ to be constant (say$ \Gamma_{0}$) then
from equation (9) ,the differential equation in H becomes

\begin{equation}
2 \dot{H} = \gamma \Gamma_{0} H - 3 \gamma H^{2},
\end{equation}

which is identical in form to that in the previous section
(equation (17)). Thus there is again scenario of emergent universe
for second order non-equilibrium thermodynamics. However, it should be noted that the present model of emergent scenario due to particle creation is an ever expanding model of the universe with constant particle creation rate and hence it has the basic features of steady state theory of Fred Hoyle et al. [25,26].

\section{Summary of The Work}

     In the present work , we have studied non-equilibrium
     thermodynamics, based on particle creation process and
     considered both first and second order deviations from
     equilibrium.Here we have concentrated on the adiabatic
     process only. In first order Eckart theory for validity of
     the second law of thermodynamics,the bulk viscous pressure
     satisfies a quadratic relation.For vanishing chemical
     potential, $\Pi$ and H are related by a simple linear
     relation and the cosmological solution obtained by solving
     the Friedmann equation gives the scenario of emergent
     universe.On the other hand ,for the second order theory the
     effective bulk viscous pressure is a dynamical variable and
     due to complicated form ,its evolution equation can not be
     solved .However, by choosing (phenomenologically) the
     constant particle creation  rate (for simplicity) and using
     the Friedmann equations ,solutions for emergent universe has
     been obtained. Thus, it is possible to have a model of
     emergent universe for both first and second order non
     equilibrium thermodynamics and therefore, it is reasonable to
     speculate that emergent scenario is a consequence of particle
     creation process and the second order model closely resembles the steady state theory.\\

     \textbf{Acknowledgement:} The work is done during a visit to IUCAA.
     The author is thankful to IUCAA for facilities at Library.
     The author is also thankful to UGC-DRS programme department
     of Mathematics,J.U.

     \textbf{References}\\

     1. E.R.Harrison,Mon.Not.R.Astron.Soc.\underline{137},69 (1967)\\

     2. G.F.R.Ellis, R.Maartens,Class.Quant.Grav.\underline{21},223(2004)\\

     3. G.F.R.Ellis,J.Murugan and C.G.Tsagas,Class.Quant.Grav.\underline{21},
        233 (2004)\\

     4. S.Mukherjee, B.C.Paul, S.D.Maharaj, A.Beesham :gr-qc /0505103 (2005)\\

     5. S.Mukherjee,B.C.Paul,N.K.Dadhich,S.D.Maharaj and A.Beesham,
        Class.Quant.Grav.\underline{23},6927 (2006)\\

     6. D.J.Mulryne ,R.Tavakol,J.E.Lidsey and G.F.R.Ellis,Phys.Rev.D\underline{71},
        123512 (2005)\\

     7. A.Banerjee, T.Bandyopadhyay and S.Chakraborty,Gravitation and Cosmology\underline{13},
        290 (2007);Gen.Relt.Grav.\underline{40},1603 (2008)\\

     8. N.J.Nunes,Phys.Rev.D\underline{72},103510 (2005)\\

     9. J.E.Lidsey and D.J.Mulryne,Phys.Rev.D\underline{73},083508(2006)\\

     10.U.Debnath, Class.Quant.Grav.\underline{25},205019 (2008)\\

     11.B.C.Paul and S.Ghose,Gen.Relt.Grav.\underline{42},795 (2010)\\

     12.U.Debnath and S.Chakraborty,Int.J.Theo.Phys.\underline{50},2892(2011)\\

     13.S.Mukerji,N.Mazumder,R.Biswas and S.Chakraborty,Int.J.Theo.Phys.\underline{50},2708(2011)\\

     14.P.Labrana,Phys.Rev.D\underline{86},083524 (2012)\\

     15.G.W.Gibbons,Nucl.Phys.B\underline{292},784 (1988)\\

     16.G.W.Gibbons,Nucl.Phys.B\underline{310},636 (1988)\\

     17.S.del Campo,R.Herrera and D.Pavan,Phys.Lett.B\underline{707},8 (2012)\\

     18.D.Pavon,S.del Campo and R.Herrera,arxiv:1212.6863(gr-qc)\\

     19.I.Prigogine,J.Geheniau,E.Gunzig and P.Nardone,Proc.Nat.Acad.Sci.U.S.A\underline{85},7428 ;Gen.Relt.Grav.\underline{21},767(1989)\\

     20.T.Harko and F.S.N.Lobo, Phys.Rev.D\underline{87},044018(2013)\\

     21.S.Saha and S.Chakraborty, in preparation\\

     22.W.Zimdahl,Phys.Rev.D\underline{53},5483 (1996)\\

     23.W.Zimdahl,Phys.Rev.D\underline{61},083511(2000)\\

     24.C.Eckart,Phys.Rev.\underline{58},919 (1940)\\
     
     25.F.Hoyle,G.Burbidge and J.V.Narlikar,Astrophysical Journal\underline{410},437(1993); Mon. Not. Roy. Astro. Soc.\underline{267},1007(1994); \underline{269},1152(1994); Proceedings of the Royal Society A\underline{448},191(1995)\\
     
     26. F.Hoyle,G.Burbidge and J.V.Narlikar,"A Different Approach to Cosmology", (2000) (Camb. Univ. Press)

\end{document}